# Influence of contact angle on slow evaporation in 2D porous media.

H.Chraïbi[1,2], M. Prat[1,2*], O.Chapuis[1,2]

[1]Université de Toulouse ; INPT, UPS ; IMFT

Avenue Camille Soula - F 31400

[2]CNRS ; IMFT - F 31400

We study numerically the influence of contact angle on slow evaporation in 2D model porous media. For sufficiently low contact angles, the drying pattern is fractal and can be predicted by a simple model combining the invasion percolation model with the computation of the diffusive transport in the gas phase. The overall drying time is minimum in this regime and is independent of contact angle over a large range of contact angles up to the beginning of a transition zone. As the contact angle increases in the transition region, cooperative smoothing mechanisms of the interface become important and the width of the liquid gas interface fingers that form during the evaporation process increases. The mean overall drying time increases in the transition region up to an upper bound which is reached at a critical contact angle $\theta_c$. The increase in the drying time in the transition region is explained in relation with the diffusional screening phenomenon associated with the Laplace equation governing the vapor transport in the gas phase. Above $\theta_c$ the drying pattern is characterized by a flat traveling front and the mean overall drying time becomes independent of the contact angle. Drying time fluctuations are studied and are found to be important below $\theta_c$, i.e. when the pattern is fractal. The fluctuations are of the same order of magnitude regardless of the value of contact angle in this range. The fluctuations are found to die out abruptly at $\theta_c$ as the liquid gas interface becomes a flat front.

---

[*] Author for correspondence : prat@imft.fr

## I. INTRODUCTION

Prediction of evaporation from porous media is of interest for many environmental and industrial applications, such as the water exchange between soil and atmosphere, drying of many products or the recovery of volatile hydrocarbons from underground oil reservoirs, to name only a few. In these applications, the liquid is generally wetting and drying can then be regarded as a drainage process, where the evaporating wetting liquid is replaced by a nonwetting gas. This case has been the subject of many studies in recent years within the framework of pore–network models and invasion percolation theory, e.g. [1], [2] and references therein. However, there are also important applications in which evaporation occurs in a hydrophobic porous medium. For example, buildings stones can be rendered hydrophobic in order to limit the salt weathering hazards due to the salt crystal formations resulting from evaporation, [3]. More generally, rendering the porous medium hydrophobic is often considered for reducing the evaporation from porous surfaces or soils. PEFCs (polymer electrolyte fuel cells) is another example in which evaporation, as well as other two-phase flow processes, takes place in partially hydrophobic (teflonized) porous (fibrous) layers. A proper understanding and modeling of two phase flows, including evaporation, in teflonized systems is needed in relation with the problem of water management of PEFCs, [4]. Yet, it is often considered that the wettability condition may change in time and this may be the cause of performance degradation. Hence it is important to study the impact of wettability change on evaporation. Here we explore the case where the wettability condition, i.e. the contact angle, is uniform throughout the porous medium. The study of situations implying spatial variations of contact angle is left for future works. As shown in [5] for the simpler process of mechanical quasi-static displacement, the invasion pattern changes from a fractal pattern to a compact one as the contact angle rises. The compact pattern is obtained for contact angles greater than a critical angle $\theta_c$. We are especially interested in the transition region right below $\theta_c$ where the invasion pattern changes significantly. We expect a similar pattern transition for the slow evaporation process considered in this paper.

Experimentally, it is not obvious to study in details the effect of contact angle on drying pattern in the transition zone since it is no easy to impose at will a given value of $\theta$. As shown in Figure 1, a partial illustration on the effect of contact angle can, however, be obtained from a simple drying experiment of a model porous made of a monolayer of 1mm glass beads randomly distributed and sandwiched between two glass plates. Without additional treatment, this leads to a contact angle $\theta \sim 30\text{-}40°$ (measured in the liquid phase). To obtain a much higher contact angle, the glass beads and the glass plates are rendered hydrophobic by silanization, a process which leads to a contact angle of the order of 105°-107°, [6]. The model is initially fully saturated by pure water. Vapor escapes through the four open lateral edges of model. The 8.8 cm long and 7.5 cm wide model is placed horizontally in a small transparent Plexiglas chamber of controlled temperature ($22 \pm 1^\circ C$ ). The relative humidity in the chamber is stabilized using a LiCl saturated solution (RH =12 %) and the evolution of phase distribution within the porous medium is recorded using a CCD camera set above the chamber. More details on this experiment can be found in [7]. For $\theta \sim 30\text{-}40°$, one can observe in Figure 1 the highly ramified capillary fingerings and trapped liquid clusters typical of invasion percolation patterns as expected. In contrast, the invasion pattern is much more compact with no trapping for $\theta \sim 105°\text{-}107°$ and resembles the faceted pattern that will be shown bellow for $\theta=99°$ (see Figure 7). In [8], we analyzed and compared the two regimes flanking the transition region. In the present effort, we concentrate on the transition region. This is made possible by taking into account in the simulations all the local mechanisms involved in the growth of the liquid-gas interface.

Throughout the text the wettability of the system is characterized by the contact angle $\theta$ measured in the dense phase, i.e. liquid water in the present context. This contact angle can be interpreted as the receding contact angle since menisci recedes during invasion of pores as a result of evaporation. As shown in a previous study, [8], the change in wettability from a hydrophilic materials ($\theta << 90°$) to a hydrophobic one ($\theta >> 90°$) leads to a major change in terms of invasion pattern and therefore also in terms of evaporation rate. As in [8], we study this change and its impact on drying rates in the quasi-static limit, i.e. when the pressure evolution in the liquid phase is only due to capillary effects. The study is conducted for a simple two dimensional (2D) model porous medium constructed by placing disks of random radii on a square lattice under the assumption of dilute concentration of water vapor in the gas phase (which is acceptable for evaporation at temperatures close to the ambient temperature). For convenience, the liquid is referred to as "water" throughout the text. It is obvious that the results are general and apply to other volatile liquids (under the condition of slow evaporation).

As shown in previous works, e.g. [8], [9], [10] and references therein, liquid films can be a major transport mechanism in the drying of porous materials. Liquid films in drying refer to "thick" films trapped by capillary effects in corners, grain contacts or surface roughness in the pores (as opposed to the thin films adsorbed by solid surface forces). For the not too small contact angles of interest in this study, it is, however, reasonable to neglect the effect of films, see [8] for more details.

The paper is organized as follows. In Sec.II the model of drying developed to simulate the evaporation process in the two dimensional (2D) model porous medium is presented. The influence of contact angle on drying is discussed in Sec. III through the analysis of evolution of mean overall drying time and drying time statistical fluctuations with the contact angle. We close in Sec. IV by offering some concluding remarks.

## II MODEL

### A. Model porous medium, transport mechanisms and external boundary condition

For simplicity, we consider a situation where heat transfer can be neglected. This corresponds typically to evaporation of a not too volatile liquid, e.g. water, at the ambient temperature. As in [5], we use a 2D array of disks with random radii (Figure 2). As shown in [5], a distinguishing advantage of this model is to allow for a full solution of the interface shape. In particular, it is possible to study in detail the influence of contact angle on the invasion pattern resulting from the evaporation process. In this model, the liquid-gas interface consists of circular arcs connecting disks. One can refer to [5] for more details on the advantage of this model compared to the more traditional pore network models made of interconnected channels. In our case, the disks are placed on a square lattice with lattice constant $a$. There are $L$ disks per row and $L$ rows on the lattice. Disk radii are distributed randomly according to an uniform distribution law in the range $[r_{min}, r_{max}]$, see details in Table I regarding the particular systems considered in this paper. The porosity $\varepsilon$ in Table I is the fraction of area not covered by disks.

Initially, the pores of the model porous medium are saturated with liquid, which is then allowed to evaporate isothermally. As depicted in Figure 2, the three upper sides of the square domain are sealed, i.e. they represent the impervious surfaces. The fourth lowermost side of the domain is open to air for drying. On this side, we use a standard mass-flux boundary condition of the form

$$j = h(P_{v,o} - P_{v,\infty}) \tag{1}$$

where $j$ is the evaporation flux density, $h$ is the mass transfer coefficient, $P_{v,o}$ is the vapor partial pressure at the porous medium surface, and $P_{v,\infty}$ is the vapor partial pressure in the surrounding air. The mass transfer coefficient $h$ is expressed as

$$h = \frac{D}{(R/M_v) T \delta} \quad (2)$$

where $D$ is the diffusion coefficient of water vapor, $R$ is the universal gas constant, $M_v$ is the vapor molecular weight, $T$ is temperature, and $\delta$ is the external transfer length-scale (we took $\delta = a$ throughout this paper). Referring to the classical convective drying situation, $\delta$ can be seen as the average thickness of the mass boundary layer developing at the porous medium surface.

At the surface of the menisci, the vapor partial pressure is the saturation vapor partial pressure $P_{vs}$ and the difference in partial pressure ($P_{vs} - P_{v,\infty}$) can be regarded as the mass transfer potential driving the evaporation process.

**B Fluid interface geometry and interface growth mechanisms**

As in [5], the interface between the gas phase and the liquid phase consists of sequence of arcs between pairs of disks. As sketched in Figure 3, each arc intersects both disks at the proper contact angle $\theta$. At the beginning of the process, we assume that the medium is saturated with liquid and the interface is formed by $L$ stable arcs attached all along the first row of disks, see Figure 2. Then as a result of evaporation, the curvature of the arcs increases until one arc becomes instable. This leads to the invasion of adjacent pore as depicted in Figure 3 and new arcs are positioned at the entrance of the new interfacial pores. Then this process is repeated until full invasion of the system. Hence, in this quasi-static limit, it is assumed that the invasion is characterized by a succession of stable configurations, with very rapid motion of the interface between two successive stable configurations.

As discussed in [5], three basic types of local instability may be distinguished:

1) "burst" when the capillary pressure between the two fluids is such that it becomes impossible to find a stable arc between two disks intercepting them with the right angle $\theta$.

2) "touch" the arc connecting two disks intersects during its growth another disk at the wrong $\theta$.

3) "overlap" = menisci coalescence, two neighboring arcs on the interface intersect.

Figure 4 shows the relative importance of various growth mechanisms as a function of contact angle $\theta$ in our system to reach breakthrough, i.e. when the invading gas phase reaches the top edge. As can be seen from Figure 4, the interface growth is dominated by bursts for sufficiently small contact angles whereas the overlap mechanism is dominant for sufficiently high contact angles. As discussed in details in [5], this change in the probabilities of each local growth mechanism is responsible for major changes in the invasion pattern, from invasion percolation patterns to a compact pattern with no trapping.

To construct the evaporation algorithm, we begin by considering the simpler case of the quasi-static mechanical displacement of the fluid in place by the invading fluid. Contrary to [5] where the interface growth was driven by small changes in the pressure applied to the system, the procedure is slightly different in our case and can be considered equivalent to imposing a very small flow rate. In particular, only one pore is invaded between two successive stable configurations of the interface. Figure 5 shows the various cases of pore invasion considered in our model. A pore occupied by the liquid phase and adjacent to a pore occupied by the invading gas phase is defined as an interfacial pore. We determine for each interfacial pore the local instability mechanism (burst, touch or overlap) that is encountered first during the growth of the interfacial arcs and denote by $R(i)$ the corresponding curvature radius of the arc(s) associated with the considered interfacial pore, i.e. the curvature radius at the onset of the local instability. The instability radii $R(i)$ are obtained by computing the evolution of the curvature radius of each interfacial arc, as it moves between the two disks it connects. The computation is analogous to that presented in [5]. For this reason, the details, which are presented in [11], are omitted here. We can then define as pore invasion potential

$$P(i) = \frac{R(i)}{a} \qquad (3)$$

which is inversely proportional to the pressure difference between the two fluids needed for the invasion of the pore ($= \frac{\gamma}{R(i)}$ where $\gamma$ is the surface tension). Under these circumstances, the invasion algorithm becomes analogous to the invasion percolation algorithm, [12], i.e. the interfacial pore of greatest potential is invaded at each step of the invasion. The fundamental difference lies in the computation of invasion potential (3), which takes into account all possible local instability mechanisms and not only the burst mechanism as in the invasion percolation algorithm. If one adds the rule that the trapped pores, i.e. the liquid pores that become completely surrounded by the invading phase, cannot be invaded, one obtains an algorithm describing the quasi-static displacement of one phase by the other under the condition of a very small flow rate of injected fluid, i.e. when the capillary forces dominate the displacement.

As first proposed in [13], the algorithm describing slow evaporation can be constructed by combining the mechanical quasi-static displacement algorithm described above and the computation of diffusive transport of the water vapor in the gas phase. This will make the process time dependent through the computation of evaporation rate, the (variable) time step being of the order of $\rho_l V_p / F$, where $\rho_l, V_p$ and $F$ are the liquid density, average volume of one pore and evaporation mass flux at the boundary of liquid cluster to which belongs the considered pore, respectively. More precisely the (quasi-)steady diffusion equation $\Delta P_v = 0.$ is solved in the part of the pore space occupied by the gas phase with the boundary conditions: $P_v = P_{vs}$ on the interfacial arcs, and Eq.(1) at the entrance of pore located along the first row of model porous medium, see [13] for more details regarding the computation of $P_v$ thanks to a finite volume type discretization technique, which has been straightforwardly adapted here to the particular case of disks array.

For the sake of completeness, the drying algorithm is summarized here: (1) every liquid cluster present in the network is identified, (2) the interfacial pore of greatest invasion potential is identified for each cluster, the volume of liquid contained at time $t$ in this pore is $V_{sc}$, (3) the evaporation flux $F_c$ at the boundary of each cluster is computed from the finite volume computation of the liquid vapor partial pressure in the gas phase (obtained from the numerical solution of equation $\Delta P_v = 0.$), (4) for each cluster, the time $t_c$ required to evaporate the amount of liquid contained in the interfacial pore identified in step (2) is computed: $t_c = \dfrac{\rho_\ell V_{sc}}{F_c}$, (5) the element among the elements selected in step (2) eventually invaded is that corresponding to $t_{cmin} = \min(t_c)$, (6) the phase distribution within the network is updated, which includes the partial evaporation of liquid contained in the interfacial pores selected in step (2) with $\rho_\ell V_\ell(t + t_{c\min}) = \rho_\ell V_\ell(t) - F_c t_{c\min}$ (where $\rho_\ell$ is the liquid density and $V_\ell$ the volume of liquid contained in the interfacial pore) except for the interfacial pore selected in step (5) which becomes completely saturated by the gas phase. The procedure can be repeated up to full drying or stopped at some intermediate stage.

**III RESULTS**

As mentioned in the introduction the two regimes limiting the transition zone were studied in [8], where it was shown that the dimensionless average overall drying time of a hydrophilic system was system size dependent approaching exponentially a limit for large size system. However, the dependence with the size is relatively weak. Furthermore, the statistical fluctuations of the drying time were found to be of the same order of magnitude for all system sizes investigated. Therefore, we can concentrate here on small systems without loss of generality and do not explore again the (weak) influence of system size. In passing, it can be noted that the computational time of drying for one realization is significantly greater here due to the consideration of all local instability mechanisms of the interface. This makes more difficult the numerical study since fewer realizations can be computed for a similar computational effort.

**A. Influence of contact angle on overall average drying time and drying pattern**

The overall drying time $t_o$ is the time needed to fully evaporate the liquid contained in the model porous medium. Figure 6 displays the evolution of overall drying time as a function of contact angle for the four systems considered. To obtain the results shown in Figure 6, we have considered water at a temperature of 20°C as working fluid and taken $a$ = 1mm. Unless otherwise mentioned, the results presented in this paper were obtained considering 100 realizations of a 25x25 disk array.

Interestingly, the average overall drying time is independent of contact angle for sufficiently low or high contact angles (as shown by the plateaus on the left hand side and the right hand side of the curves in Figure 6). These plateaus correspond to the regimes studied in [8], i.e. the invasion percolation regime for sufficiently low contact angle when the growth of the interface takes place in the largest local constriction available along the interface of a liquid cluster (this regime corresponds to bursts as dominant local instability mechanisms) or, for sufficiently high contact angle, when the pattern is characterized by a flat traveling front (menisci overlaps are then the dominant local instability mechanism). These two regimes are termed IP (invasion percolation) regime and FF (flat invasion front) regime in the following. The results shown in Figure 6 for the IP and FF regimes are consistent with the ones obtained in [8] for a pore network model of interconnected channels of rectangular cross-section, i.e. drying in the hydrophilic model (IP regime) is faster than in the hydrophobic one (FF regime). The ratio between the lowest and largest average drying times is slightly greater than 0.6 as shown in the inset of Figure 6 for the disk array whereas a ratio slightly higher of 0.7 was obtained with the pore network model in ref. [8] for a network of comparable size. Interestingly, Figure 6 shows that the transition between the IP and FF drying times occurs over a relatively narrow range of contact angles. As shown in Figures 4 and 6, the greater the porosity, the wider the transition zone.

In the analysis of the pressure driven quasi-static invasion process considered in [5], a critical contact angle $\theta_c$ was introduced to characterize the transition between uniform flooding and fractal invasion patterns. Noting that the contact angle is defined in the invading phase in [5] and not in the invaded phase as in our case, see Figure 2, the results reported in [5] indicates that uniform flooding occurs above $\theta_c$ (using our definition for the contact angle). As it will be made clear from the consideration of drying patterns and drying time fluctuations, $\theta_c$ corresponds to the contact angle marking the end of transition region on the right hand side in Figure 6, i.e. the contact angle marking the beginning of the FF regime. Approximate values of $\theta_c$ deduced from Figure 6 are reported in Table I for four porosities. Consistently with the results of [5], $\theta_c$ appears to increase monotically with porosity. Whereas $\theta_c$ increases with porosity, the contact angle marking the beginning of the transition region is found to be independent of porosity and close to 90°, as can be seen from Figure 6.

It is interesting to observe that the simple drying model proposed in [13], which is based on a the IP algorithm for modeling the capillary effects, is sufficient to predict the average overall drying time over the full range of contact angle [0, 90°], provided that liquid film effects can be neglected, see [9] and references therein for more details on the effect of films. Referring to both Figures 4 and 6, it can be seen that the average overall drying time begins to be affected only when the probability of overlap mechanism becomes greater than the burst mechanism probability.

Figure 7 shows examples of drying patterns for system $D$ ($\varepsilon = 0.7$). The same realization of the 25x25 disk array is considered for four different values of the contact angle, namely $\theta=80°$ (invasion percolation (IP) regime), $\theta=94°$ (transition region), $\theta=99°$ (transition region) and $\theta=120°$ (flat traveling front = FF regime). The IP regime ($\theta=80°$) is characterized by ramified fractal patterns and the occurrence of many disconnected liquid clusters whereas the FF regime ($\theta=120°$ in Figure 7) is characterized by an almost flat traveling invasion front with no trapping. As shown in [5] for the case of the quasi-static fluid invasion and illustrated in Figure 7 for the case of the evaporation problem considered here, the trapping phenomenon progressively disappears as the contact angle increases in the transition region. The growth of the interface is increasingly smoother as the contact angle is increased in the transition region and this is due to the increasingly cooperative nature of invasion as the overlap mechanism becomes the dominant local instability mechanism. As explored in detail in [5], larger and larger segments of the interface move forward coherently as the contact angle increases in the transition region and this leads to the "faceted" interface shown in Figure 7 for $\theta=99°$. For a sufficiently high contact angle (above $\theta_c$ with the convention for the contact angle used here), the entire interface advances coherently and this leads to the FF regime.

The fact that the average drying time increases with the contact angle in the transition region can be understood, at least qualitatively, from the evolution of the drying patterns shown in Figure 7 and the results reported in [5]. If one ignores the external mass transfer resistance (~ $h^{-1}$ from Eq.(1)), which is independent of the contact angle, the evaporation flux roughly varies as $d^{-1}$ where $d$ is the average distance between the evaporation front inside the porous medium and the exposed surface (bottom side in Figure 7) of the porous medium. Hence the deeper the evaporation front inside the porous medium, the lower the evaporation rate. When the invasion pattern is flat and compact ($\theta=120°$ in Figure 7), the evaporation front coincides with the flat invasion interface. For a given number of invaded pores (or equivalently a given overall liquid saturation), this corresponds to the deepest evaporation front position that is possible to obtain since there is no "holes" in the domain occupied by the remaining liquid. Hence this type of invasion pattern leads to the largest drying time. In the other cases (transition region and IP regime), the evaporation front does not coincide anymore with the interface between the two fluids and this leads to faster drying time. Note that the equation governing the transport of water vapor in the gas phase is the Laplace equation $\Delta P_v = 0$. It is well known that the transport governed by the Laplace equation is characterized by the phenomenon of diffusional screening when it takes place from an irregular interface, e.g. [14] and references therein. Here this means that the local evaporation flux density is not uniform along the interface but strongly varies from one place to another. For example in the case of the IP patterns shown in Figure 7 ($\theta=80°$), the evaporation flux density is quasi-null at the boundary of the fjords located deep inside the porous medium due to diffusional screening and the evaporation flux density is in fact only significant at the menisci located along the outermost boundary of the liquid / gas region shown in Figure 7.

A simple manner of illustrating the diffusional screening effect is to shown the invasion pattern together with the vapor partial pressure field $P_v$. This is shown in Figure 7. As expected the most marked screening effect is obtained for the IP regime where the interface is the most ramified and the numerous liquid clusters, close to each other, contribute to the screening phenomenon. From the vapor isoconcentration lines shown in Figure 7, it is obvious that the "effective" evaporation front is much less deep inside the porous model for the IP regime than for the FF regime (for the same overall saturation in both systems). Hence the faster drying observed in the IP regime is a consequence of the diffusional screening effect together with the fact that bursts are the local dominant instability mechanisms in this regime (which lead to the IP highly ramified structure). When the growth of the interface becomes more coherent, the screening effect is less effective as the ramified fjords of the IP regime transform into large gulfs. This is illustrated in Figure 7 ($\theta = 99°$). As a result, the effective evaporation front is located deeper in the porous medium on the average and this leads to greater average overall drying time compared to the IP regime. This can be expressed in terms of the average finger width $\overline{w}$ studied in [5]. The finger width can be interpreted here as the mean distance between two menisci. In the IP regime, $\overline{w}$ is on the order of pore size and the screening effect is very effective. In the transition region, $\overline{w}$ (the size of gulfs) increases (up to divergence at $\theta_c$) with $\theta$ and the screening effect becomes less and less effective.

This can be characterized quantitatively through the consideration of the equivalent flat evaporation front, which can be defined as follows. Suppose that $j_{\theta c}(t)$ is the evaporation flux density at time $t$ for the FF regime, i.e. for $\theta \geq \theta_c$. $j_{\theta c}(t)$ can be expressed as

$$j_{\theta c}(t) = \frac{M_v}{RT} \left( \frac{\delta}{D_{ext}} + \frac{d_{\theta c}(t)}{D_{pm}} \right)^{-1} (P_{vs} - P_{v,\infty}) \tag{4}$$

where $d_{\theta c}$, $D_{ext}$ and $D_{pm}$ are the position of the flat front (= distance between the front and the porous medium surface), the effective diffusion coefficient in the external boundary layer and the porous medium effective diffusion coefficient respectively. Defining the external mass transfer resistance as $G_e = \dfrac{\delta}{D_{ext}}$, it can be seen from Eq. (4) that $j_{\theta c}(t)$ scales as

$\left(G_e + \dfrac{d_{\theta c}(t)}{D_{pm}}\right)^{-1}$. Hence from the numerical data for $\theta \geq \theta_c$, we can plot $j_{\theta c}^{-1}$ as a function of

$d_{\theta c}(t)$ and determine the values of $G_e$ and $D_{pm}$. Then, according to Eq.(4), the equivalent flat front position for any flux density $j_\theta$ can be defined as

$$d_{eq} = D_{mp}\left(\dfrac{(P_{vs}-P_{v,\infty})}{j_\theta}\dfrac{M_v}{RT} - G_e\right) \quad (5)$$

Figure 8 shows the evolution of $d_{eq}(\theta)/d_{\theta c}$ (averaged over 100 realizations) as a function of the overall mean saturation (the overall saturation is the volume fraction of the pore space occupied by the liquid). As can be seen from Figure 8, the diffusional screening is very effective for $\theta < \theta_c$, which explains the faster drying as discussed above

To conclude this section, it can be observed that the influence of contact angle on average drying time can be still more marked when an intermediate low overall saturation is sought. This is shown in Figure 9 where $\langle t_S \rangle_\theta$ is the average drying time needed to reach the mean overall saturation $<S>$ when the contact angle is $\theta$. As can be seen from Figure 9, the evolution of the relative difference in drying time as a function of mean overall saturation is not monotonic. The maximum difference is reached at a low saturation close to 0.5 for our computations. The decrease in the drying difference for lower saturations is associated with the drying of the last row of pores, see [8] for some details on this very last period of drying.

## B. Influence of contact angle on statistical fluctuations

As shown in [8], the drying time is subject to significant statistical fluctuations in the IP regime whereas the drying time variability was nearly null for the FF regime, which is obvious from the invasion pattern. More precisely, it was found that the distribution of overall drying time was nearly Gaussian with $\sigma / \langle t_0 \rangle_{IP} \approx 0.1$ for a 25 x 25 network, where σ is the standard deviation of overall drying time over many realizations. Hence it is interesting to study how the magnitude of statistical fluctuations varies in the transition zone. Figure 10 shows the evolutions of $\sigma / \langle t_0 \rangle$ in the transition zone. As can be seen the ratio $\sigma / \langle t_0 \rangle$ increases in the transition zone up to the sudden decrease when the invasion pattern corresponds to the FF regime, i.e. when there is no more spatial fluctuations in the growth of the interface. Hence it is remarkable to observe that the statistical fluctuations of the drying time are important up to $\theta_c$, which is consistent with the existence of random invasion patterns below $\theta_c$. The behavior of $\sigma / \langle t_0 \rangle$ near $\theta_c$ is shown in Figure 11 and is consistent with a linear variation of the form $\sigma/\langle t_0 \rangle \propto (1 - \theta/\theta_c)$. Figure 12 shows the evolution of PDF (probability density function) of overall drying time in the transition region. As mentioned earlier, the computational time of one realization is significantly greater than for the simpler drying models used in [5]. As a result, the PDFs shown in Figure 12 were obtained considering 10000 realizations of a 15x15 model porous medium and not 100000 as in [8] (recalling that only the IP regime was considered in [8]). Despite the more limited number of realization considered here, a clear evolution of the overall drying time PDF is observed through the transition region. For $\theta \geq \theta_c$, the drying time fluctuations die out and the PDF is a dirac delta function (not shown in Figure 12). For the IP regime ($\theta = 80°$ in Figure 12), the PDF is consistent with the nearly Gaussian distribution obtained in [8] for a 2D square network of interconnect channels. As noted in [8], the drying time distribution is in fact slightly dissymmetric and skewed to the left for the IP regime. As can be seen from Figure 12, this dissymmetry of the PDF significantly increases with the contact angle in the transition region. This can be also shown from the values of the skewness reported in Table II. Note the tail on the left hand side for $\theta = 99°$ and the complete dissymmetry for $\theta = 102°$, i.e. as $\theta_c$ is approached from below. Although probably difficult to obtain experimentally, the dissymmetry of the drying time PDF is a clear signature of the transition zone, i.e. of the changes in the local interface growth mechanisms.

These results are for the overall drying time and it can be surmised that the effect of statistical fluctuations can be still greater for intermediate average drying time. This is illustrated in Figure 13 which shows, for several values of contact angle, the evolution of the standard deviation $\sigma_s$ of the overall liquid saturation over 100 realisations of a 25 x 25 disks array as a function of the average saturation $\langle S \rangle$ over the 100 realizations, whereas the insert in Figure 13 shows how $\langle S \rangle$ varies with time $t$. As can be seen from Figure 13, the maximum magnitude of $\sigma_s$ is similar for all values of contact angle below $\theta_c$ and close to 0.08. However, the average saturation at which this maximum is observed increases with the contact angle.

It can be concluded that the drying time fluctuations are significant below $\theta_c$ and die out at $\theta_c$. However, one should recall here that the liquid films have been ignored in the analysis. As stated in the introduction, it is reasonable to neglect the effect of films for the relatively high contact angles of interest in this study, see [8] for more details. Interestingly, the results presented in [8] about the influence of liquid film in the IP regime indicates that the statistical fluctuations are strongly dampen when the films can develop over a significant distance toward the surface of porous medium ahead of the liquid cluster region. Although the details remain to be elucidated, this effect is due to an enhanced screening effect of the heterogeneity of liquid phase distribution in the liquid cluster region.

This suggests that the drying time statistical fluctuations and related fluctuations are significant only over a limited range of contact angles. The upper bound of this range is well defined and is given by $\theta_c$. The lower bound is less well defined since the results presented in [8] (noting in passing that the "critical" angle $\theta_c$ used in [8] should not confused with the angle $\theta_c$ considered in the present study) suggests that the damping of fluctuations due to films is not abrupt but takes place over a (albeit narrow) range of contact angle.

## IV CONCLUSIONS

In this paper, we have studied the influence of contact angle on slow evaporation in 2D model porous media. For sufficiently low contact angles, the drying pattern is fractal and can be predicted by a simple model combining the invasion percolation model with the computation of the diffusive transport in the gas phase. The overall drying time is minimum in this regime and is independent of contact angle over a large range up to the beginning of a transition zone. For the cases considered in this study, the transition zone spans over about 10° in contact angle variation. As the contact angle increases in the transition region, cooperative smoothing mechanisms of the interface become important and the width of the liquid gas interface gulfs that form during the evaporation process increases. As a result, the diffusional screening phenomenon becomes less effective and the mean overall drying time increases up to an upper bound which is reached at a critical contact angle $\theta_c$. Above $\theta_c$ the drying pattern is characterized by a flat traveling front and the mean overall drying time becomes independent of the contact angle. Drying is found to be almost 65% slower for contact angles above $\theta_c$ compared to the minimum drying time observed for contact angles below the contact angle marking the beginning of the transition region (~ 90° in our case).

Below $\theta_c$ the drying time fluctuations induced by the spatial fluctuations in the liquid phase distribution associated with each realization are significant. The drying time distribution standard deviation is of the same order of magnitude regardless of the value of contact angle in this range. However, there is a significant change in the drying time PDF. The PDF is nearly Gaussian at the beginning of the transition zone and becomes more and more skewed to the left as the contact angle increases in the transition zone. The fluctuations die out abruptly at $\theta_c$ as the liquid gas interface becomes a flat front.

Liquid films were neglected in the analysis. From previous studies, e.g. [8],[9] it is expected that films do not affect significantly the invasion patterns but do affect the mean drying time and drying time fluctuations. However, this should no affect the analysis of the transition region presented in this study since film effects are expected to be non negligible only for sufficiently low contact angles, i.e. for contact angles expected to be significantly lower than $\theta_c$.

Our study was restricted to a square lattice of discs with random radii. Disks on a triangular lattice were also considered in [5] and the variations of invasion pattern with $\theta_c$ exhibited the same behavior as the square lattice. Thus, evaporation in triangular lattices as well as in other similar random systems is expected to exhibit the same trends as the system considered in the present study. This is illustrated in part by the experimental results shown in Figure 1 as well as those reported in [15] for a 2D network of channels of rectangular cross section.

As pointed out in [5], a transition region is expected to be observed also in 3D. According to [5], the critical contact angle $\theta_c$ is, however, expected to be lower in 3D. The main difficulty with a 3D model is that an accurate description of the evolution of the liquid-gas interface shape becomes much more involved. Despite this difficulty, it is certainly desirable to consider the 3D case, which is obviously more representative of real porous media, since it is well known that there are significant differences in the structure of the phase distribution during drying (for contact angle below $\theta_c$) compared to the 2D case owing to the fact that each phase can form a percolating cluster in 3D, e.g. [16] . To develop a 3D pore network model a evaporation for any value of contact angle, one option is to use the approximate parametric models of pore capillary entry pressure developed in previous works on two-phase flows , e.g. [17]. Work in this direction is in progress.

Also, only capillary forces were considered in the study and the increase in the drying time in the transition region and for $\theta \geq \theta_c$ is entirely associated with the change in the drying pattern due to the change in the local instability mechanisms controlling the growth of the interface. Under other circumstances, in particular when the drying pattern is not only controlled by the capillary forces but by the competition between capillary forces and gravity or viscous forces, the increase in drying time with the contact angle can be due to a completely different mechanism, i.e. a change in the invasion pattern due to the reduction in capillary forces compared to gravity or viscous forces (and therefore not to a change in the interface local growth mechanisms). As discussed for instance in [18], it is well know that gravity forces as well as viscous forces can contribute to stabilize the drying patterns, i.e. to the formation of a drying front. This should be kept in mind when analyzing experimental data, especially when $\theta_c$ is close to 90°. For example, the change in the saturation profiles observed in the drying experiments analyzed in [19] between a random packing of 250μm beads with receding contact angles of 0° and a similar packing but with a receding contact angle of 84° (referred to as "hydrophobic" beads in [19]) could be interpreted at first glance as consistent with the transition from IP pattern to flat traveling front analyzed in the present study. However, as pointed out in [19], the influence of gravity effects is important, compared to capillary forces, with the "hydrophobic" beads. We note that the front is not sharp in this experiment with the "hydrophobic" beads and the saturation profiles are in fact quite consistent with a 3D version of the gravity stabilized front discussed in [18]. Hence, it can be surmised that $\theta_c >$ 84° for the system studied in [19]. Naturally, both effects, i.e. the change in the local growth mechanisms and the change in the competition between capillary forces and gravity or viscous forces can be responsible for the increase of drying time with the contact angle in real systems.

**ACKNOWLEDGEMENTS**

Financial support from GIP ANR : (project ANR-06-BLAN-0119-01 « Intensifilm ») is gratefully acknowledged. We thank N.Sghaier for permission to present the experimental patterns shown in Figure 1.

[1] A.G. Yiotis, A.K., Stubos, A.G. Boudouvis, I.N., Tsimpanogiannis, Y.C. Yortsos, Transport in Porous Media **58**, 63 (2005). A.G. Yiotis, I.N. Tsimpanogiannis, A.K. Stubos and Y.C. Yortsos J. Colloid Interface Science **297**, 738 (2006). T.Metzger, E.Tsotsas and M.Prat, *Modern Drying Technology*, (Wiley, New-York 2007), Vol.1, Chap.2.

[2] M.Prat, Chem. Eng. J., **86**, 153 (2002).

[3] A. Goudie, H.Viles, *Salt weathering hazards*, (Wiley, Chichester, 1997).

[4] C.Y. Wang, in Handbook of Fuel Cells, W.Vielstich, H.A. Gasteiger, A.Lamm (Eds), Vol.3, Part 3, p.337 , Wiley, 2003.

[5] M.Cieplak, M.O.Robbins, Phys. Rev. B. **41**, 16, 11508 (1990), N.Martys, M.Cieplak, M.O.Robbins, Phys. Rev. Lett. **66**,8, 1058 (1991).

[6] N. Sghaier, M. Prat, S. Ben Nasrallah, Chem.Eng. J., **122**, 47–53, (2006)**.**

[7] N.Sghaier, M.Prat submitted to Transp. In Porous Media

[8] O.Chapuis, M.Prat, Phys. Rev. E, **75**, 046311 (1-11)(2007)

[9] M.Prat, Int.J. of Heat and Mass Tr. , **50**,1455-1468 (2007)

[10] J.B.Laurindo and M. Prat, Chem. Eng. Sci., **53** (12), 2257 (1998).

[11] O.Chapuis, INPT, 2006, Ph-D thesis (in French), unpublished,

[12] D.Wilkinson, and J.F.Willemsen, J.Phys.A-Math.Gen.,**16**,3365 (1983)

[13] M.Prat, Int. J. Mult. Flow, **19**, 691 (1993)

[14] D.S.Grebenkov, M.Filoche, B.Sapoval, Phys.Rev. E, **73**, 021103 (2006)

[15] O. Chapuis, M. Prat, M. Quintard, E. Chane-Kane, O. Guillot, N. Mayer, J. of Power Sources, **178**, 258 (2008)

[16] Y.Le Bray, M. Prat M., Int. J. of Heat and Mass Tr.,**42**,4207,(1999)

[17] M.J.Blunt, J.Pet.Sci. Eng., **20**, 117, (1998). P.H.Valvatne, M.J.Blunt, Water Resour.Res., **40**, W07406 (2004)

[18] M.Prat, F.Bouleux, Phys. Rev.E, **60,**5647 (1999)

[19] N.Shahidzadeh-Bonn, A.Azouni, P.Coussot, J.Phys. :Condens. Matter, **19**, 112101 (2007)

List of Tables and Figures

Table I Parameters of the system studied (the lattice spacing *a* is used as reference length scale)

| System | Range of radii | Porosity | $\theta_c$ (°) |
|---|---|---|---|
| A | 0.32-0.349 | 0.63 | 98 |
| B | 0.29-0.349 | 0.65 | 99 |
| C | 0.27-0.349 | 0.67 | 100 |
| D | 0.23-0.35 | 0.70 | 101 |

Table II Standard deviation ($\sigma$), skewness (*sk*) and kurtosis ($\beta$) for the drying time distributions shown in Figure 12.

| Contact angle $\theta$ (°) | $\sigma/\langle t_0(\theta) \rangle$ | *sk* | $\beta$ |
|---|---|---|---|
| 80 | 0.145 | -0.243 | 2.515 |
| 94 | 0.166 | -0.339 | 2.556 |
| 99 | 0.179 | -0.705 | 2.947 |
| 102 | 0.100 | -2.487 | 10.31 |

Figure 1 Example of phase distribution during drying in a model porous medium made of a monolayer of 1 mm glass beads randomly distributed between two glass plates. Vapor escapes from the four sides of model porous medium: a) $\theta \sim$ 30-40° b) $\theta \sim$ 105°-107°. The black areas correspond to liquid saturated regions. Glass beads forming the porous medium are visible in the regions (in grey) invaded by the gas phase.

Figure 2 Model porous medium formed by a regular array of disks of random diameter. Distribution of fluids at *t = 0* (invading fluid in grey, displaced fluid in white). The interface is visible as a series of arcs joining the bottom row of disks. Vapor escapes from bottom edge of system

Figure 3 Schematic example of arcs location before and after invasion of one pore and definition of contact angle.

Figure 4 Probability (in %) of interface local growth mechanisms as a function of contact angle in the displaced fluid.

Figure 5 Growth of interfacial arcs during the invasion. The arrows indicate the meniscus displacement direction. Local invasion events: a) burst, b) touch, c) and d) arc coalescence (overlap)

Figure 6 Mean overall drying time $\langle t_0 \rangle$ as a function of contact angle; $t_{ref} = t_{oD}(\theta_c)$, i.e. the overall drying for system D for $\theta \geq \theta_c$. The inset shows the evolution of $\langle t_0 \rangle / t_o(\theta_c)$, where $t_o(\theta_c)$ is the overall drying for the considered system for $\theta \geq \theta_c$.

Figure 7 Example of drying patterns (liquid phase in dark gray, gas phase in light gray) in the transition region for one realization of a 50x50 disks array (system D). Value of contact angle

is indicated at top of each vertical series of patterns; the numbers of invaded pores from top row to bottom row are 1250, 1500, 1750 and 2000, respectively . The 5 vapor isoconcentration lines shown together with each pattern correspond to $P_v/P_{vs}$ =0.2 (the closest to the porous medium surface) , 0.4, 0.6, 0.8 and 0.9995 (the farthest from the porous medium surface), respectively.

Figure 8 Equivalent flat front position as a function of mean overall saturation for various values of contact angle.

Figure 9 Evolution of $\dfrac{t_S(\theta_c) - \langle t_S \rangle_\theta}{t_o(\theta_c)}$ as a function of average overall saturation. $t_0(\theta_c)$ is the overall drying time for $\theta = \theta_c$. $\langle t_S \rangle_\theta$ and $t_S(\theta_c)$ are the mean drying time needed to reach a given average saturation $S$ for $\theta = \theta_c$ and the drying time to reach $S$ when $\theta = \theta_c$, respectively.

Figure 10 Standard deviation σ of overall drying time as a function of contact angle in the transition region for different porosity

Figure 11 Evolution of overall drying time standard deviation near $\theta_c$ (~101°). The inset shows the variation of σ over the transition zone

Figure 12 PDF (probability density function) of overall drying time for four values of contact angle (corresponding to symbols on the curve in the inset, which shows the evolution of mean overall drying time as a function of contact angle in the transition region for 15x15 disk systems).

Figure 13 Evolution of standard deviation of overall liquid saturation $\sigma_S$ as a function of average saturation $\langle S \rangle$. The inset shows the evolution of $\langle S \rangle$ as a function of time.

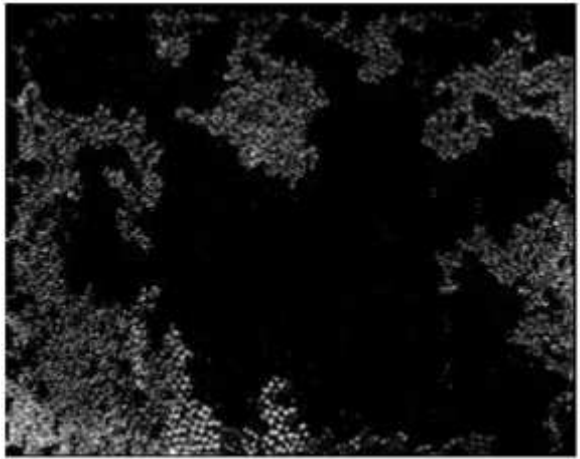 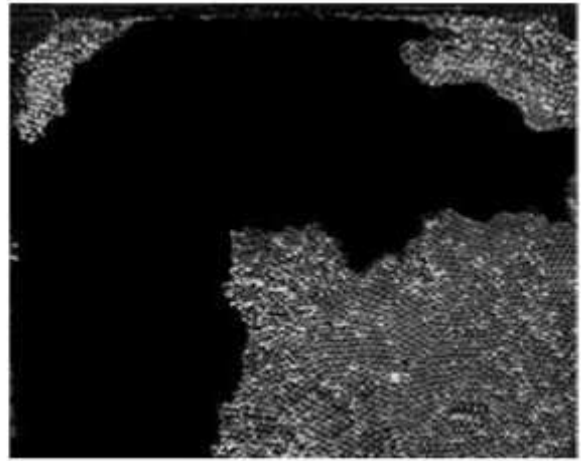

a) b)

Figure 1

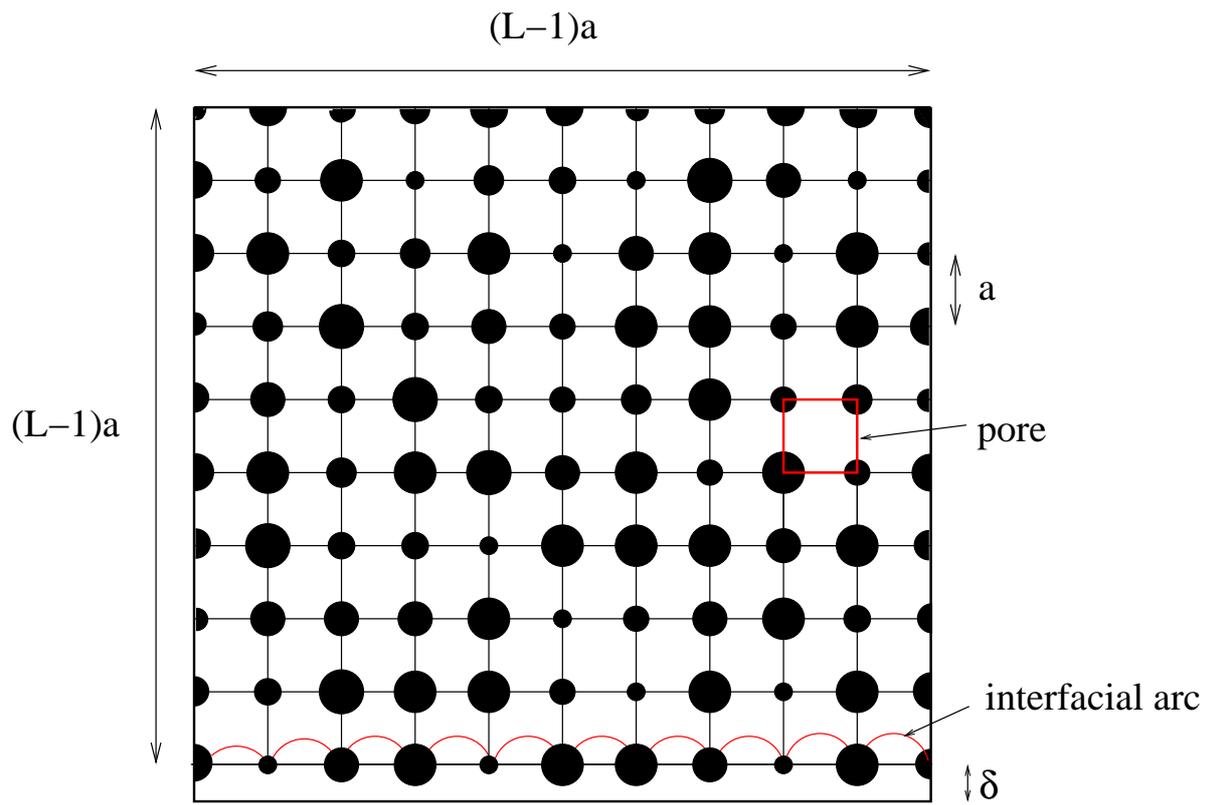

Figure 2

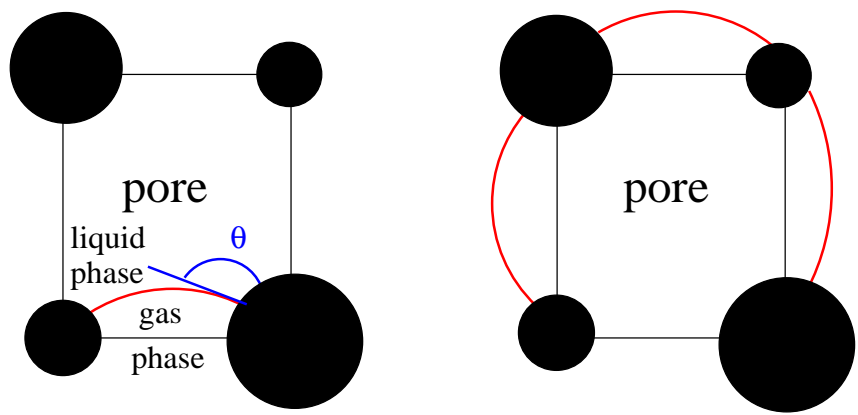

Figure 3

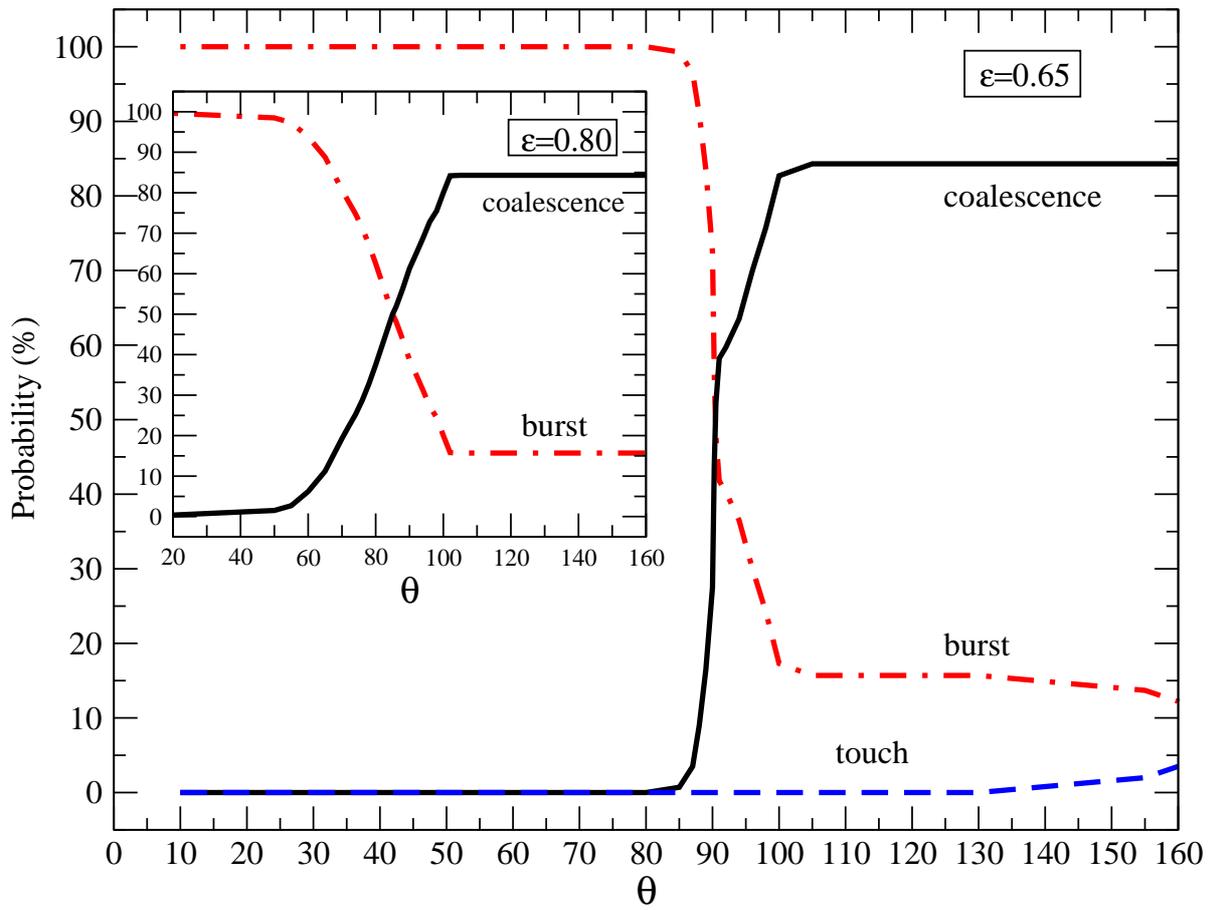

Figure 4

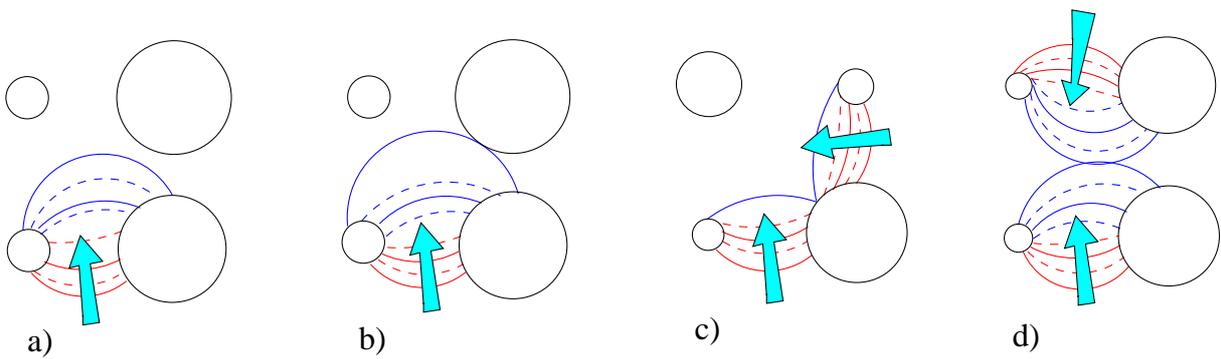

Figure 5

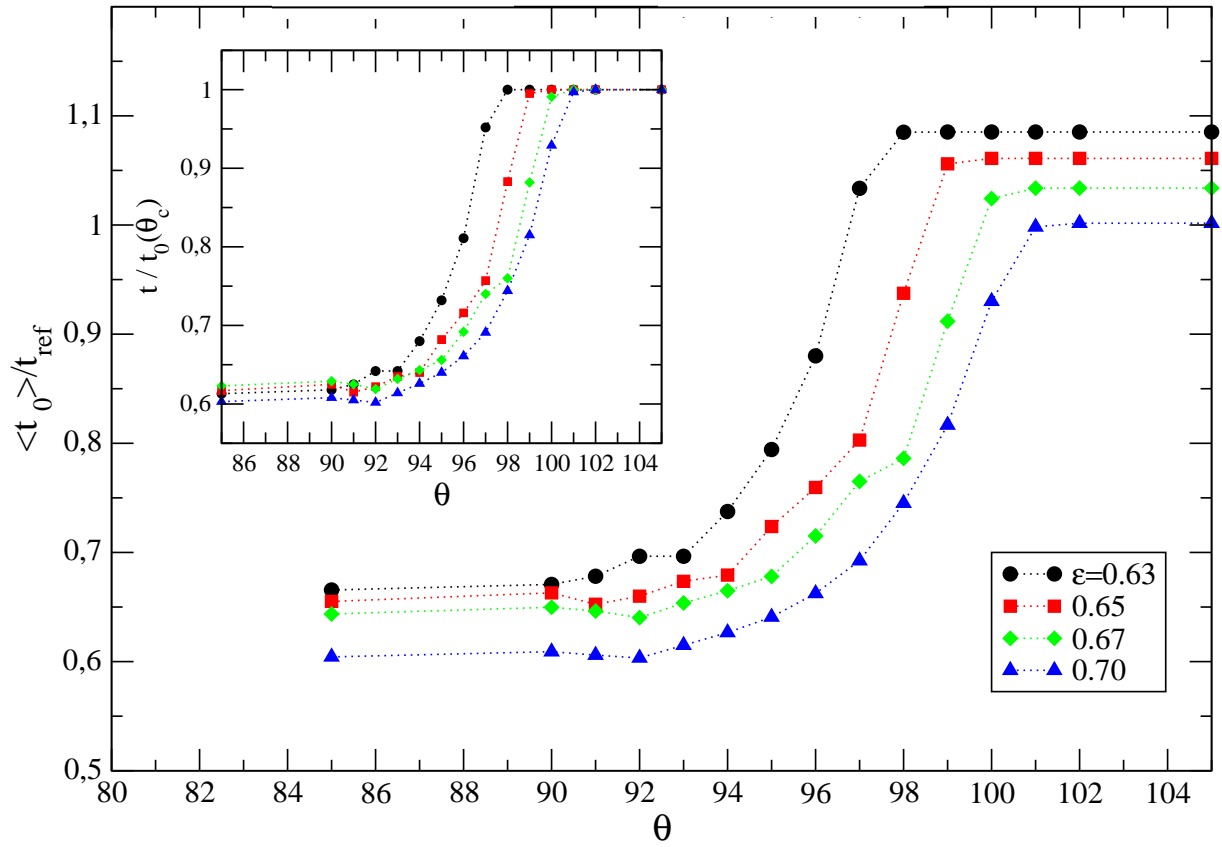

Figure 6.

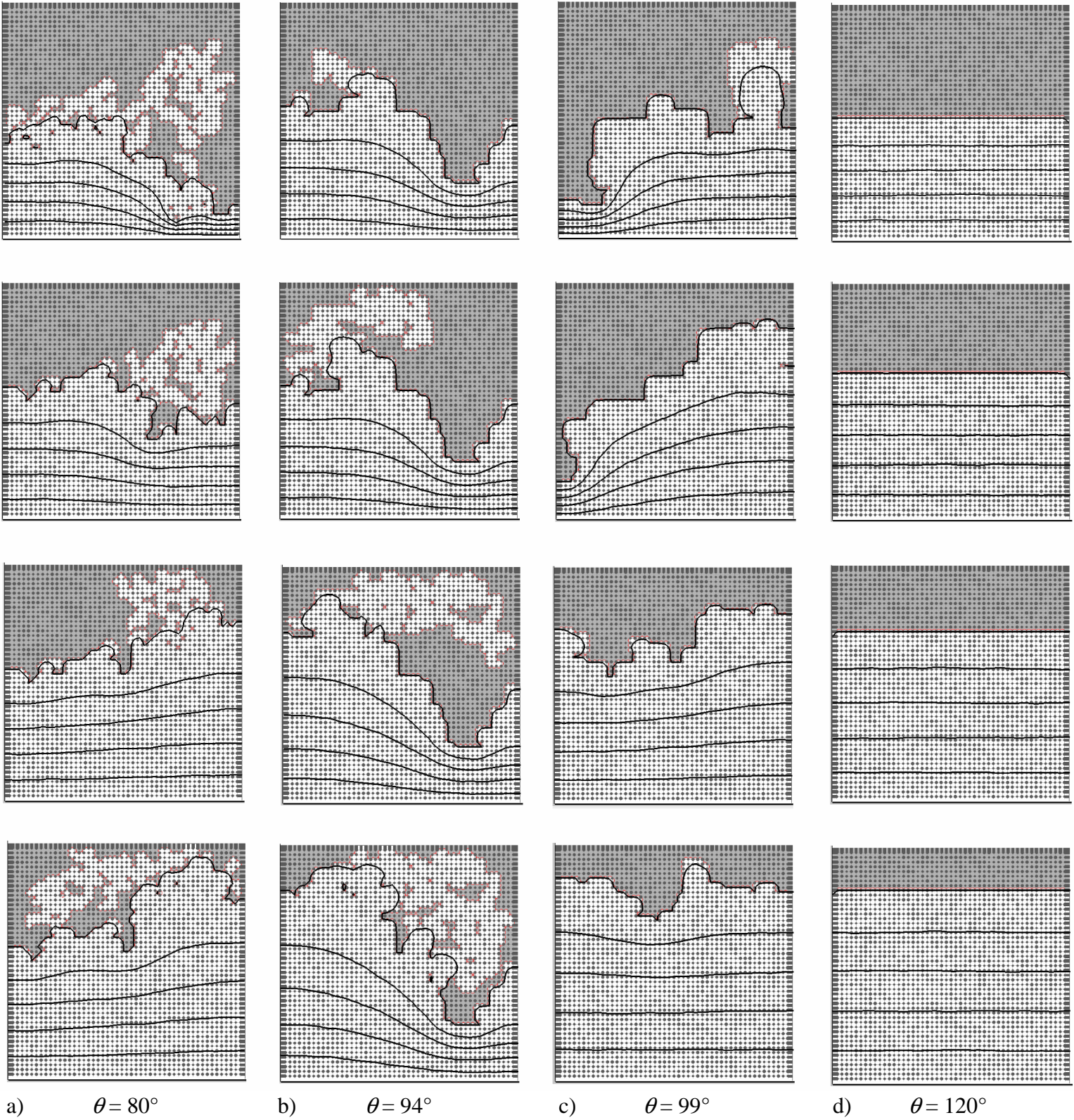

a)  θ = 80°     b)  θ = 94°     c)  θ = 99°     d)  θ = 120°

Figure 7

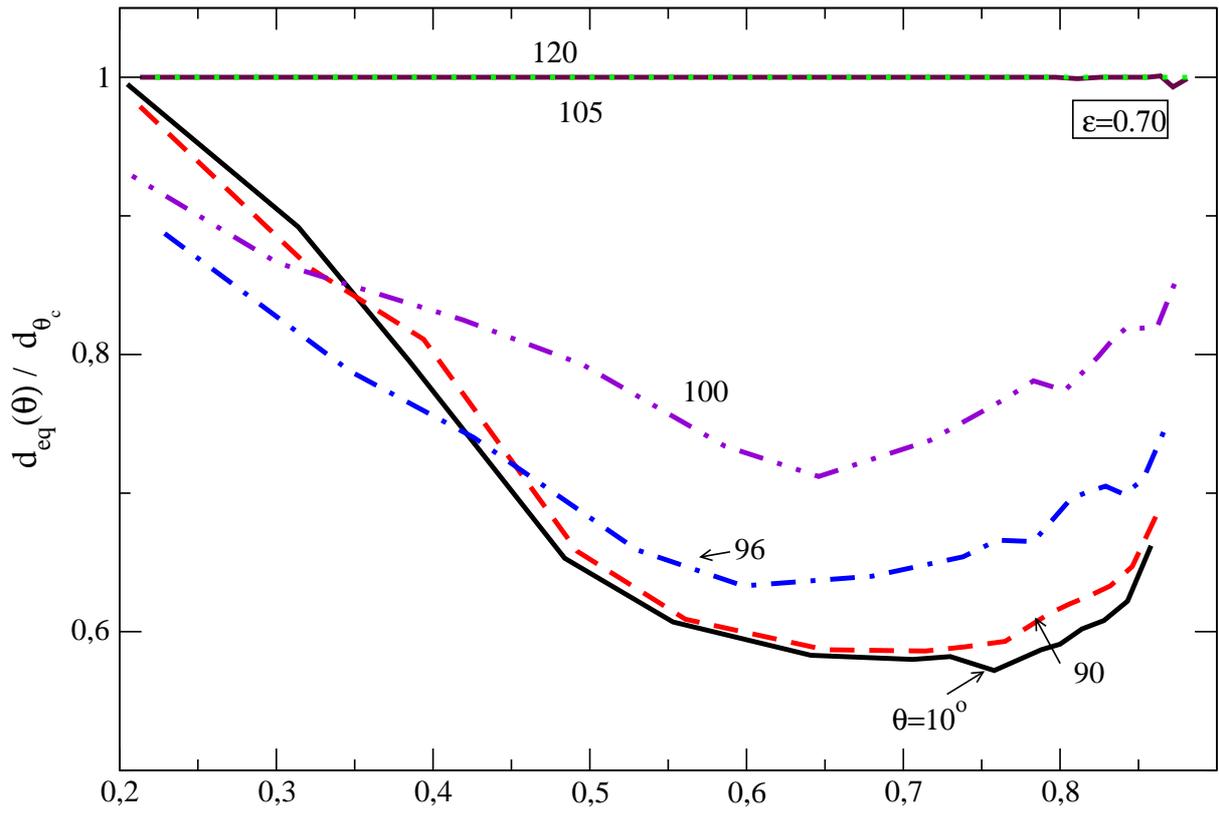

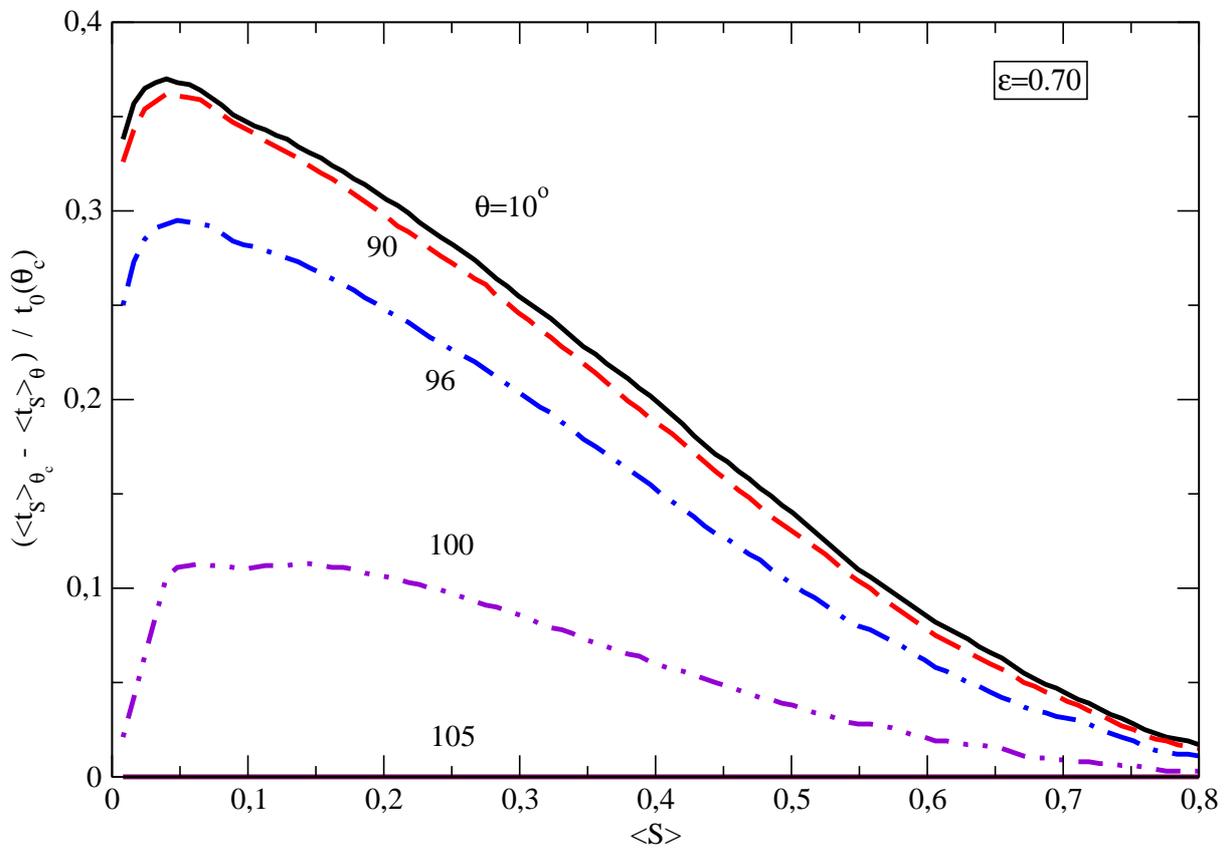

Figure 9

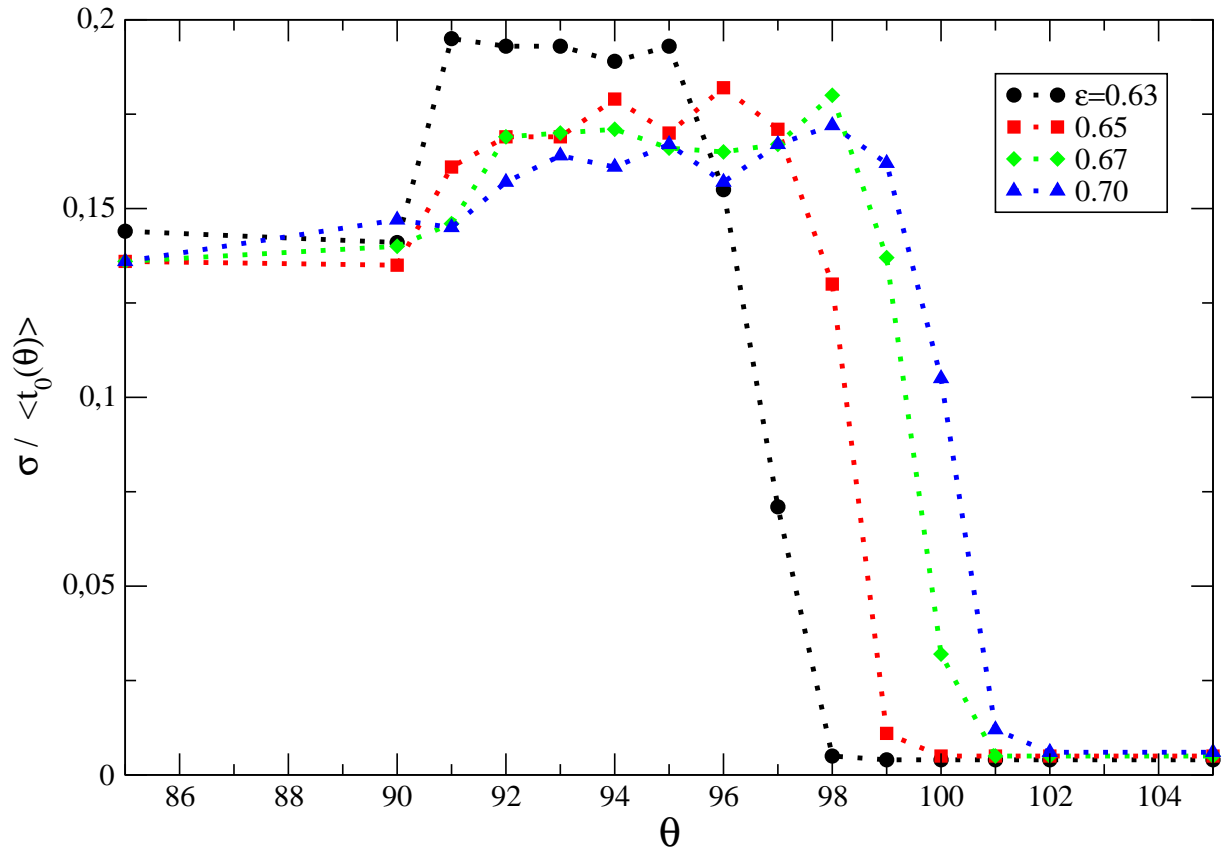

Figure 10

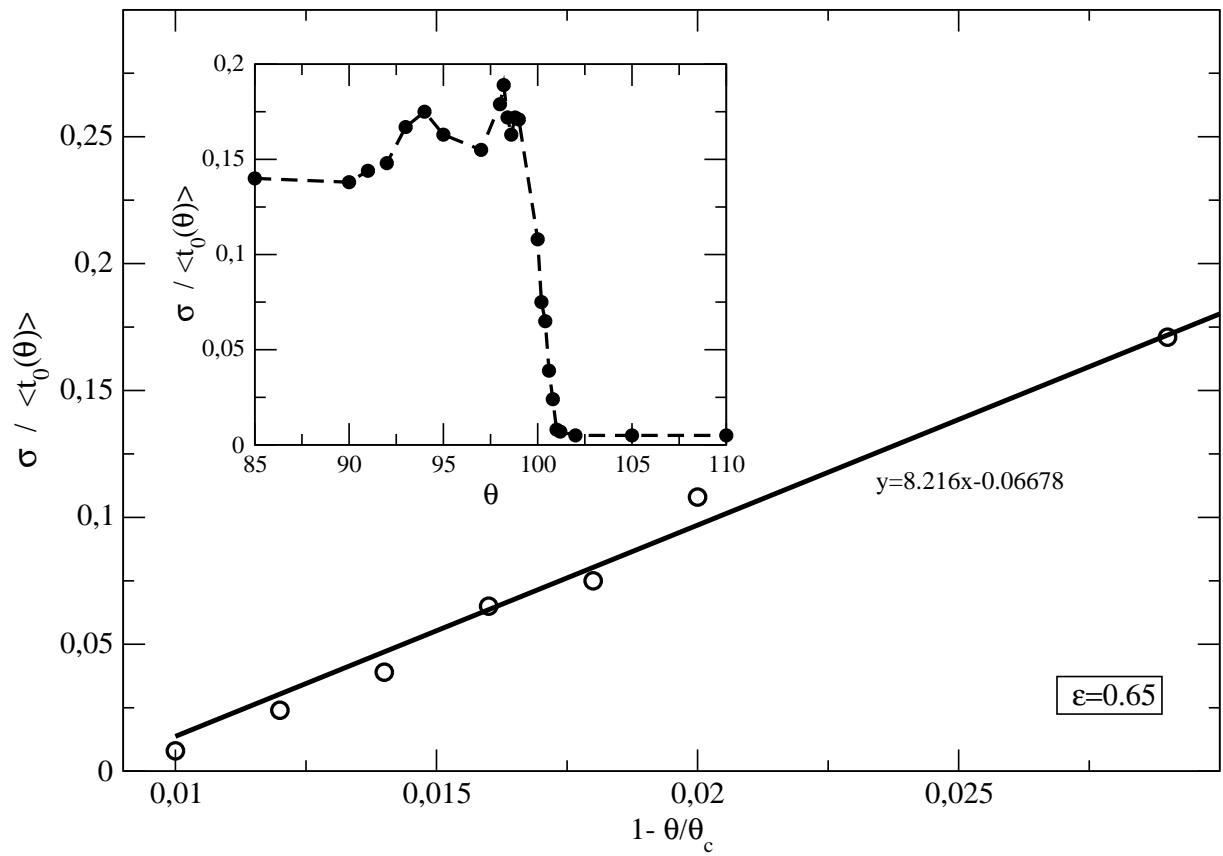

Figure 11.

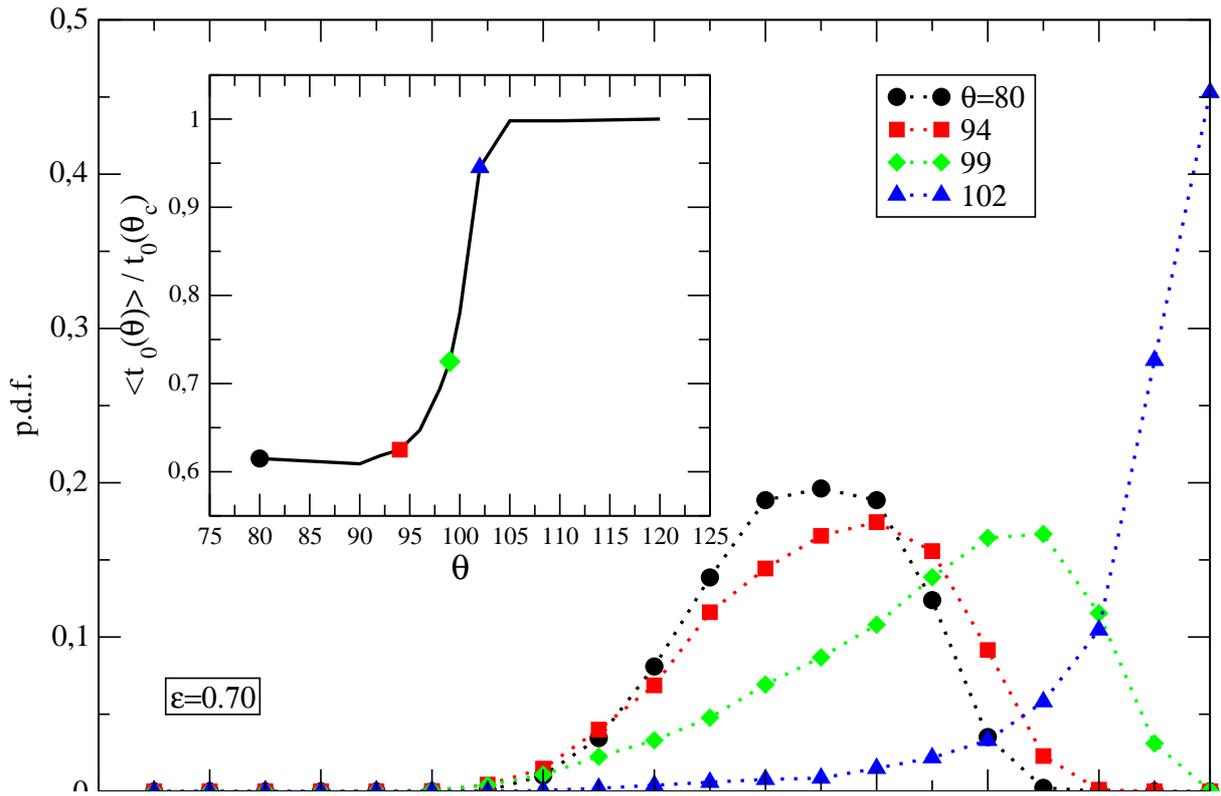
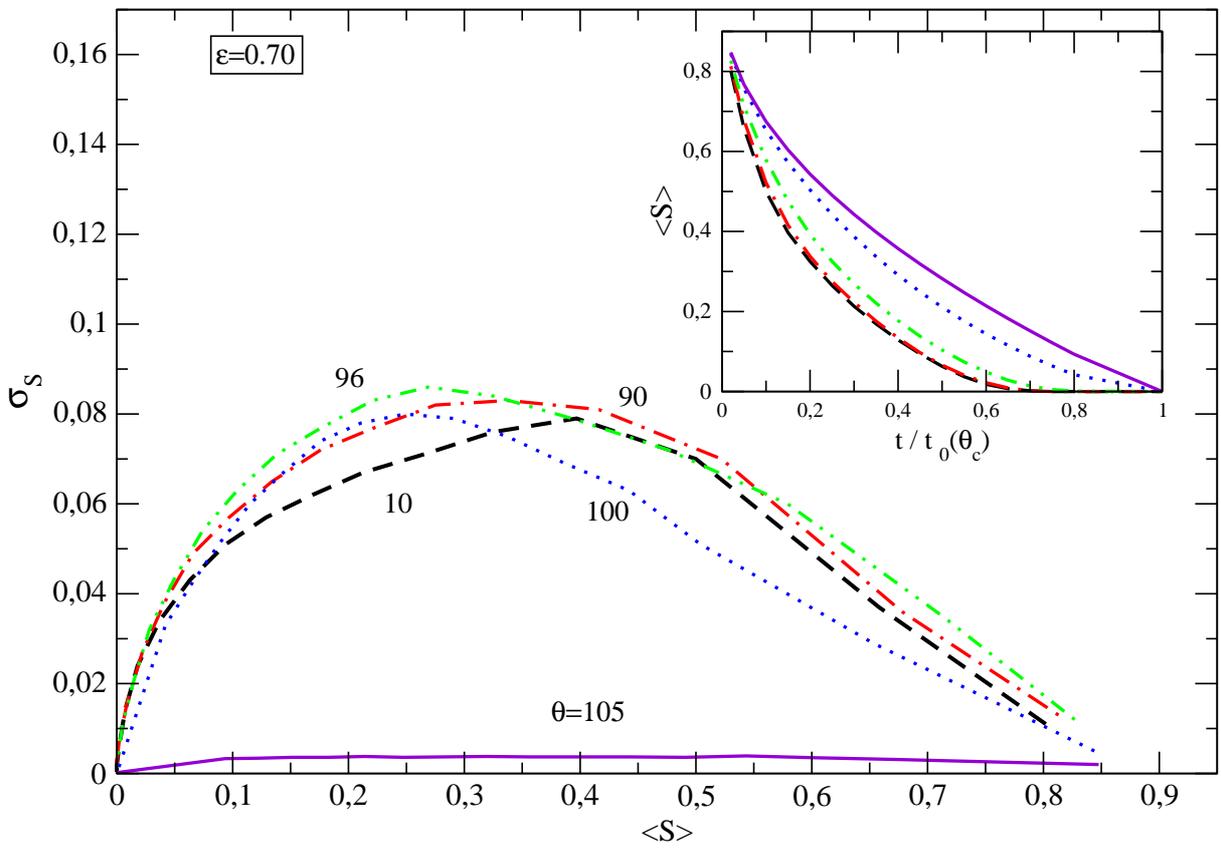

Figure 13